\RequirePackage{fix-cm}
\documentclass[smallextended]{svjour3}       
\smartqed  
\usepackage{graphicx}
\usepackage{lineno,hyperref}
\usepackage{amsmath,amssymb,array}
\usepackage{booktabs}

\usepackage{caption}
\usepackage{subcaption}
\usepackage{footnote}

\usepackage[dvipsnames]{xcolor}

\usepackage{footnote}
\makesavenoteenv{tabular}
\makesavenoteenv{table}

\usepackage{float}

\usepackage{natbib}
\setcitestyle{sort,numbers,square}
\begin{document}
	
	\title{An empirical estimation for time and memory algorithm complexities: Newly developed R package
	}
	
	
	\author{Marc Agenis-Nevers         \and
		Neeraj Dhanraj Bokde       \and
		Zaher Mundher Yaseen       \and
		Mayur Shende
	}
	
	
	\institute{M. Agenis-Nevers \at
		Independent Researcher, Epicentre Factory, Clermont-Ferrand, France\\
		\email{marc.agenis@gmail.com}           
		\and
		N. D. Bokde \at
		Department of Engineering - Renewable Energy and Thermodynamics,	Aarhus University, Denmark \\
		\email{neerajdhanraj@eng.au.dk}
		\and
		z. M. Yaseen \at
		Sustainable Developments in Civil Engineering Research Group, Faculty of Civil Engineering, Ton Duc Thang University, Ho Chi Minh City, Vietnam \\
		\email{yaseen@tdtu.edu.vn}
		\and
		M. Shende \at
		Government College of Engineering, Nagpur, India
	}
	
	\date{Received: date / Accepted: date}

	\maketitle
	
	\begin{abstract}
	{When an algorithm or a program runs on a computer, it requires some resources. The complexity of an algorithm is the measure of the resources, for some input. These complexities are usually space and time. The subject of the empirical computational complexity has been studied in the research. This article introduces \texttt{GuessCompx} which is an R package that performs an empirical estimation on the time and memory complexities of an algorithm or a function, and provides a reliable, convenient and simple procedure for estimation process. It tests multiple increasing-sizes samples of the user's data and attempts to fit one of seven complexity functions: \emph{O(N),	O(N\^{}2), O(log(N))}, etc. In addition, based on the best fit procedure using leave one out-mean squared error (LOO-MSE), it predicts the full computation time and memory usage on the whole dataset. 
			Together with this results, a plot and a significance test are returned. Complexity is assessed with regard to the user's actual dataset through its size (and no other
			parameter). This article provides several examples demonstrating several cases (e.g., distance function, time series and custom function) and optimal parameters tuning.}
	\end{abstract}
	
	\section{Introduction}\label{introduction}
	Complexity of an algorithm is a measure of resources amount the algorithm needs to run the completion, for a given input \cite{he2001drift}\cite{jensen2003reducing}. Usually, the resources of an algorithm require either space or time, which represent two types of complexities \cite{Z1} \cite{woeginger2004space}. The time complexity is the total amount of time required by an algorithm to complete its execution \cite{Z2}. {The running time of an algorithm may differ for the same input sizes, depending on the type of data \cite{valiant1979complexity, r1a, r1b}. For instance, in bubble sort algorithm, if the input array is already sorted then the running time of the algorithm will be lower \cite{Z3}. This is what is so called the best case complexity. Similarly; if the input list is in reverse order, the algorithm will take the maximum time capacity, which is called worst case complexity. If the input list is neither sorted nor reversed, then the running time is less than worst case but more than best case: this is the average case complexity.} Since an algorithm cannot require more time than the worst case scenario, this scenario is set to be the reference complexity \cite{min2010analysis}\cite{yang2011experimental}. Since this function is generally difficult to compute exactly, and the running time for small inputs is usually not consequential, one commonly focuses on the behavior of the complexity when the input size increases. This is the asymptotic behavior of the complexity. Hence, the time complexity is commonly expressed using big \texttt{O} notation, typically \emph{O(N),
		O(NlogN), O(N\^{}2), O(2\^{}N)}, etc., where \texttt{N} is the input size in units of bits needed to represent the input \cite{chivers2015introduction}.

	{Most mathematical problem can be solved with different algorithms \cite{Z4}.
		These algorithms may include complex optimization \cite{r2a,paliwal2020computer, r2b}, forecasting \cite{r2e, r2d}, data decomposition \cite{r2c} or searching methodologies \cite{r2f}.
		These algorithms do not need the same amount of resources and can have different complexities. The algorithm that uses less memory and completes in less time is, by definition, more efficient. By studying the complexities of different algorithms, it can be determined the most efficient algorithm for a given input.}
	
	In the specific field of data science, it has been witnessed an exponential growth of datasets size and required computational power \cite{r1c, r1d}. Some review researches such as \cite{qiu2016survey} gave an overview of the statistical methods for big data, mentions the words ``computational complexity'' or ``computational efficiency'' no less than 10 times. However, while designing new algorithms, their combinations, or even simply data treatments or data streams, the engineers have a constant need to check the computational complexity of their code \cite{salih2019new}. A great amount of time can be lost in this process, while waiting for a new computation to complete without knowing the exact time required: one minute, one hour, several weeks? Is it even worth waiting? Having a way to estimate the execution time of a new piece of code, before running it in full scale, could be a critical time-saving tool.
	
	{In recent years, the possibility of new and efficient computational models including empirical views of complexity theory is recommended as a part of computer science \cite{n2}. The importance of this view and its consequences are discussed in detail in \cite{n1}. Further, the need of the empirical methods for complexity computation is convinced by the comment: ``\emph{In particular, the existence of a function computable by a machine but not by an algorithm would imply that some mathematical problem are solvable by empirical processes without any mathematical method}'' \cite{n1,n2,n3}. With this motivations, there are a couple of tools to estimate computational complexity empirically as discussed in the next paragraph along with their limitations and drawbacks. This paper has proposed \texttt{GuessCompx}, a new R package to estimate the computational complexity accurately with minimum efforts. The subsequent sections are dedicated to description and demonstration of the package.}
	
	According to \cite{kumarsharma2018predicting}, ``\emph{it takes significant amount of effort to judge the complexity of an algorithm. Various manual methods are used till now to get calculate the time complexities such as Master Method, using control flow graph, but all of them remain tedious to work with and owing to their manual nature, have limitations and are prone to error}''. Recently, manual and exact methods have started to be challenged by so-called empirical methods, that try to give an estimate of the complexity by observing the code being run several times. Goldsmith et al. \cite{goldsmith2007measuring} proposed a tool in C/C++ language named Trend Profiler (\texttt{trend-prof}) to construct models of empirical computational complexity that estimate how many times each basic block in a program runs; statistical modeling is applied, in the form of linear, polynomial or power terms, the models for the different blocks of code are eventually combined to predict the final execution time or to report performance problems to the developer. The authors don’t directly measure CPU-time and justify this choice by the desire to avoid the replicability problem. Modeling code-blocks is more robust but has a major drawback: it needs to access the source code of the target function and to ask some manual-input features like the number of nodes, characters, lines, etc. More recently, \cite{kumarsharma2018predicting} introduced a concept of predicting time complexity using gradient boosted trees in a supervised manner in C++ language and Python. The authors reached exceptional prediction capability by treating the problem not as a regression anymore but as a classification into 7 complexity families. The algorithm has to be executed in a controlled environment with a high number of replications, to retrieve 6 features linked with execution time and hardware information (clock time, number of processors, etc.), but the technique also gave good results using only two of them: the execution time and processor speed. Unfortunately, the code and the models are not directly applicable (e.g., package or library) for a data scientist to use in his daily work. Eventually, apart from those two works, the state of the art concerning memory complexity estimation seems to be particularly sparse.
	
	This research is devoted on the proposition of \texttt{GuessCompx} package \cite{Guesscompx}, which enables the R users to empirically estimate the computation time and memory usage of any algorithm prior to the final implementation. As per author’s knowledge, this is the first package proposal in CRAN which discusses empirical estimation of algorithm complexity.
	
	In the \texttt{GuessCompx} package, the complexity estimation is only defined with regard to \texttt{N} (being the number of rows of the data). Many other factors could be thought of that influence complexity, such as: data dimension, number of features, time horizon for recursive forecasting, sparsity of the data, possibility of parallel computing, and eventually interactions between two or more of those factors. Such influences are out of scope of the \texttt{GuessCompx} package, but could be investigated in future versions. Details on the subject of algorithmic complexity can be found at \cite{wiki}.
	
	The popular complexity functions have implemented over the literature are
	\emph{O(1), O(N), O(N\^{}2), O(N\^{}3), O(N\^{}0.5), O(log(N)),
		O(N*log(N))}. Those functions are representative of the most commonly found complexity functions and were chosen to be similar as those tested in \cite{kumarsharma2018predicting}, excepted for 
	\emph{O(N*N\^{}0.5)} which was replaced by the constant function. 
	
	{
		It is worth to highlight that the salient features of the proposed \texttt{GuessCompx} package in comparison with the existed techniques are:}
	
	{
		\begin{itemize}
			\item The \texttt{GuessCompx} is a user-friendly tool to estimate the empirical complexity of functions. The complexities of algorithm or functions are achievable with a single step of code and minimum efforts.
			\item The complexity accuracy with \texttt{GuessCompex} can be very high and it is examined for some popular algorithms with known complexities in the subsequent section. The accuracy can further be improved with fine-tuning the several input parameters used in the package, such as \texttt{max.time}, \texttt{start.size}, \texttt{replicates}, etc.
			\item Apart from accuracy, another major advantage of \texttt{ GuessCompx} than other tools/packages/techniques, it does not require having the code of the target function.
			\item The \texttt{GuessCompx} provides a reliable, convenient, and simple procedure for the estimation process.
			\item The \texttt{GuessCompx} is the only tool of its kind in the R and can be easily integrated with other programming languages such as Java, Matlab with several relevant R packages.	
	\end{itemize}}

	{The rest of the manuscript is organized as follows: Section 2 describes the \texttt{GuessCompx} package and its functions in further detail. Section 3 demonstrates the usefulness of the \texttt{GuessCompx} package with several examples and discusses how the best tuning parameters attained. Section 4 shows the performance of the \texttt{GuessCompx} package in predicting the time and space complexities. Finally, the summary of the proposed \texttt{GuessCompx} package is presented in Section 5.}

	\hypertarget{guesscompx-package-description}{%
		\section{\texorpdfstring{\texttt{GuessCompx} package
				description}{GuessCompx package description}}\label{guesscompx-package-description}}
	
	An asymptotic complexity behavior can be defined for most common algorithms: some are independent of the length of the data (think of the \texttt{length} function), some linear, some quadratic (typically a
	distance computation), etc. We track the computation time and memory of the algorithm runs on increasing subsets of the data, using sampling or stratified sampling if needed. We fit the various complexity
	functions with a simple \texttt{glm()} procedure with a formula of the kind \texttt{glm(time\ \textasciitilde{}\ log(nb\_rows))}, then find
	which is the best fit to the data. This comparison between the models is achieved through a LOO (\textbf{leave-one-out}) routine using Mean
	Squared Error as the indicator.
	
	The \texttt{GuessCompx} package has a single entry point: the \texttt{CompEst()} function that accept diverse input formats (data.frame, matrix, time series) and is fully configurable to fit most
	use cases: which size of data to start at, how much time the user has to do the audit (usually 1 minute gives a good result), how many replicates you needed for each tested size (in case of high variability), is a stratified sampling required (in case each run must include all possible categories of one variable), by how much we increase the size at each run, etc.
	
\begin{figure}[H]
		\includegraphics[width=\textwidth]{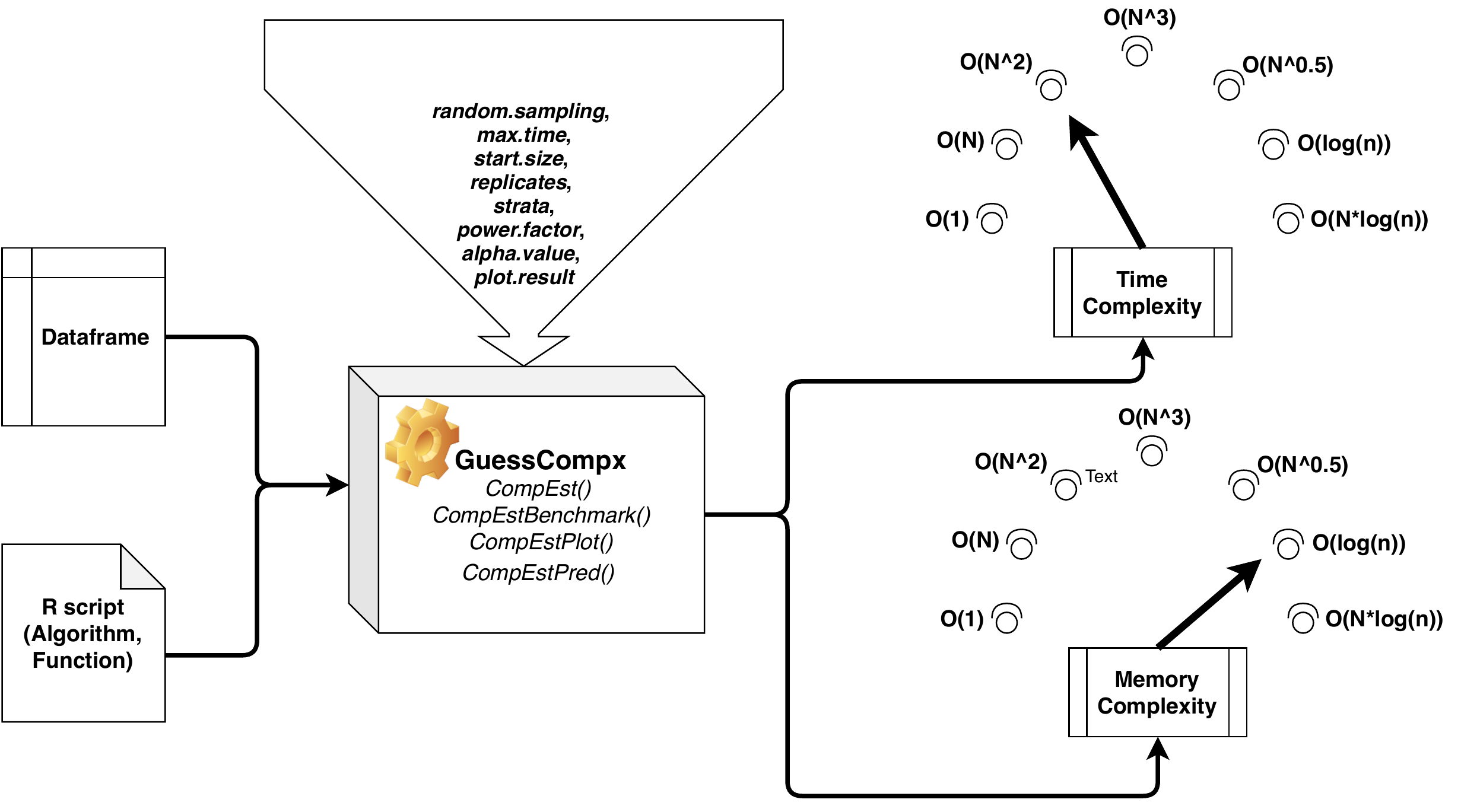} \caption{{generalized block diagram of \texttt{GuessCompx} package.}}\label{block}
	\end{figure}

	The \textbf{plot output} helps to compare the fit of each complexity function to the data. {The generalized block diagram of the \texttt{GuessCompx} package is as shown in Figure \ref{block}.}
	
	{The \texttt{GuessCompx} package is adopted to estimate the worst case complexity of an algorithm or function based on increasing length of the targeted dataset. Hence, the dataframe and the R script (or function or code line/s) of the targeted algorithm are the mandatory inputs for the \texttt{GuessCompx} package. Then after, the \texttt{GuessCompx} package can be fine tuned with several input parameters as shown in Figure \ref{block}. These parameters are discussed in detail in next subsections. Based on the inputs and tuning of the package, it provides the time and memory complexity for the given algorithm with the targeted dataframes. There are several other modes of output results and visualization (which are not mentioned in the block diagram) discussed in next section while demonstrating the \texttt{GuessCompx} package.}
	
	Note on the memory complexity: memory analysis relies on the \texttt{memory.size()} function to estimate the trend and this function only works on Windows machines. If another OS is detected, the algorithm
	will skip the memory part.
	
	\hypertarget{imports}{%
		\subsection{Imports}\label{imports}}
	
	\texttt{GuessCompx} requires the following CRAN packages to be
	installed: \textbf{\texttt{dplyr}, \texttt{reshape2},
		\texttt{lubridate}, \texttt{ggplot2}, \texttt{boot}}
	
	\hypertarget{installation}{%
		\subsection{Installation}\label{installation}}
	
	The package can be downloaded from the CRAN repositories:
	
	\footnotesize
	\begin{verbatim}
	\begin{Sinput}
	# install.packages("GuessCompx")
	library(GuessCompx)
	# ?CompEst
	\end{Sinput}
	\end{verbatim}
	\normalsize
	
	Also, it can be download it from Github:
	
	\footnotesize
	\begin{verbatim}
	# install.packages("devtools")
	library(devtools)
	install_github("agenis/GuessCompx")
	\end{verbatim}
	\normalsize
	
	\hypertarget{introduction-to-functions-in-guesscompx-package}{%
		\subsection{\texorpdfstring{Introduction to functions in
				\texttt{GuessCompx}
				package}{Introduction to functions in GuessCompx package}}\label{intoduction-to-functions-in-guesscompx-package}}
	
	Following sub-sections provide the insight on the main and internal functions
	developed in the \texttt{GuessCompx} package.
	
	\hypertarget{the-compest-function}{%
		\subsubsection{\texorpdfstring{The \texttt{CompEst()}
				function:}{The CompEst() function:}}\label{the-compest-function}}
	
	\texttt{CompEst()} is the main function used for estimation of complexity of an algorithm. The function first creates a vector of data sizes, with possible replicates, and loops through it. The \texttt{max.time} argument is then used as a stopping condition to the loop (it stops as soon as the previous iteration has exceeded the time limit). Indeed, the \texttt{max.time} argument does not limit the total time spent by the function, but the maximum time of one single iteration  including possible replicates: the total number of iterations will be high if the power factor is low, if the starting
	size is also low, and if the running time increases slowly. The user might have to think about how to set his parameters to prevent the computation time to be too long.
	
	The core of the \texttt{CompEst} function works by simultaneously evaluating the time and the memory used by one iteration of the target function over a sample of the data. Time evaluation is achieved via a
	call to \texttt{system.time()}, memory evaluation is done through \texttt{memory.size()} function, called before and after a double garbage collection \texttt{gc()}. By passing the limitations of the
	\texttt{memory.size()} function to only Windows users have been unsuccessfully tried (\texttt{pryr::mem\_change,\ Rprof}). Note that the
	sampling phase is conducted outside the time/memory evaluation, so it has no footprint on the results.
	
	A dataframe containing the data sizes, the memory and time results, is created and serves as input to fit 2*7 complexity models and add their predictions to the data. The \texttt{cv.glm()} function eventually
	computes the Leave-One-Out RMSE error of each model, in an efficient way. Eventually, once the best model is selected (the one with the smallest LOO-error), a significance test is performed to check if the model
	is better than an intercept-only model.
	
	It must be noted that, when a CONSTANT relationship is predicted, it might simply mean that the \texttt{max.time} value is too low to show
	any tendency. Several such cases rise an alert to the user to suggest modifications in the value of the function's arguments.
	
	The function returns a list with the best complexity model and the computation time on the whole dataset, for both time and memory complexity (Windows) and time complexity only (all other OS). Also, it returns two plots, with the best model curve highlighted in bold. The function has following arguments:
	
	\footnotesize
	\begin{verbatim}
	CompEst(d, f, random.sampling = FALSE, max.time = 30,
	start.size = NULL, replicates = 4, strata = NULL,
	power.factor = 2, alpha.value = 0.005, plot.result = TRUE)
	\end{verbatim}
	\normalsize
	
	\texttt{d}: A dataframe on which the algorithm is to be tested. This can
	also be a vector or a matrix.
	
	\texttt{f}: A user-defined function that runs the algorithm. The algorithm takes \texttt{d} as the first argument. There is no need for the function to return any value.
	
	\texttt{random.sampling}: This argument can have only two possible values, either TRUE or FALSE (boolean). The default value is set to FALSE. If the value is TRUE, a random sample is taken at each step. If
	FALSE, the first N observations are taken at each step. Choosing a random sampling is relevant with the use of replicates to help the discrimination power for complexity functions.
	
	\texttt{max.time}: The maximum time allowed for each step of the analysis in seconds, that is the time for all replicates of a single sample size. Once the specified time limit is reached, the execution of the function is stopped. If no value is specified, the default value of 30 seconds is used. There is no such limitation regarding memory.
	
	\texttt{start.size}: The size of the first sample to run the algorithm. The size is given in the form of rows number. The default value of the argument is \texttt{floor(log2(nrow(d)))}. If strata is not NULL, it is recommended to pass a value which is multiple of number of categories.
	
	\texttt{replicates}: The number of replicated runs of the algorithm for a specific sample size. This argument allows better discrimination of the complexity function. The default value of the argument is set to 2.
	
	\texttt{strata}: This argument is a string containing the name of categorical column of d that must be used for stratified sampling. A fixed proportion of the categories is sampled, always keeping at least one observation per category.
	
	\texttt{power.factor}: The common ratio of the geometric progression of the sample sizes. The default value is 2. It means that sample sizes double every step. The argument can also be passed as a decimal value.
	
	\texttt{alpha.value}: The alpha risk of the test whether the model is significantly different from a constant relation. The default value is set to 0.005.
	
	\texttt{plot.result}: A boolean value to indicate if the summary plot of all the complexity functions is to be displayed. The best model is shown by the curve in bold. If FALSE, then the plot is not displayed. The default value is set to TRUE.
	
	\hypertarget{the-compestbenchmark-function}{%
		\subsubsection{\texorpdfstring{The \texttt{CompEstBenchmark()}
				function:}{The CompEstBenchmark() function:}}\label{the-compestbenchmark-function}}
	
	\texttt{CompEstBenchmark()} function presents a benchmark procedure to fit complexity functions to a dataframe of time or memory values. The function takes input a dataframe produced by the \texttt{CompEst()}
	function. User needs to specify whether function deals with time or memory data. The function returns a list of all the fitted complexity model.
	
	The function has following arguments:
	
	\footnotesize
	\begin{verbatim}
	CompEstBenchmark(to.model, use = "time")
	\end{verbatim}
	\normalsize
	
	\texttt{to.model}: A dataframe produced by the \texttt{CompEst()}
	function. The dataframe  is comprised of size, time, memory and
	\emph{NlogN\_X} columns.
	
	\texttt{use}: A string indicating if the function deals with time or memory data. The default value of the argument is ''time''.
	
	\hypertarget{the-compestplot-function}{%
		\subsubsection{\texorpdfstring{The \texttt{CompEstPlot()}
				function:}{The CompEstPlot() function:}}\label{the-compestplot-function}}
	
	The \texttt{CompEstPlot()} function plots the results of algorithms
	complexities. It returns a ''ggplot'' object.
	
	The arguments of the function are:
	
	\footnotesize
	\begin{verbatim}
	CompEstPlot(to.plot, element_title = list("", ""), use = "time")
	\end{verbatim}
	\normalsize
	
	\texttt{to.plot}: A dataframe produced by \texttt{CompEst()} function.
	
	\texttt{element\_title}: A string that will be added to the subtitle of
	the plot.
	
	\texttt{use}: A string to indicate if the function deals with ''time''
	or ''memory'' data. The default value is ''time''.
	
	\hypertarget{the-compestpred-function}{%
		\subsubsection{\texorpdfstring{The \texttt{CompEstPred()}
				function:}{The CompEstPred() function:}}\label{the-compestpred-function}}
	
	The \texttt{CompEstPred()} function predicts the computation time of the whole dataset. The function returns the predicted time for the whole dataset.
	
	The arguments are:
	
	\footnotesize
	\begin{verbatim}
	CompEstPred(model.list, benchmark, N, use = "time")
	\end{verbatim}
	\normalsize
	
	\texttt{model.list}: A list containing the fitted complexity functions,
	produced by \texttt{CompEst()} function.
	
	\texttt{benchmark}: A vector of LOO errors of complexity functions,
	produced by \texttt{CompEst()}.
	
	\texttt{N}: Number of rows of the whole dataset, produced by
	\texttt{CompEst()}.
	
	\texttt{use}: A string indicating if the function deals ''time'' or
	''memory'' data.
	
	\hypertarget{groupedsamplefracatleastonesample-and-rhead}{%
		\subsubsection{\texorpdfstring{\texttt{GroupedSampleFracAtLeastOneSample()}
				and
				\texttt{rhead()}}{GroupedSampleFracAtLeastOneSample() and rhead()}}\label{groupedsamplefracatleastonesample-and-rhead}}
	
	The \texttt{GroupedSampleFracAtLeastOneSample()} function samples a random proportion of data, keeping at least one observation. This function is designed to allow its use with group splitting or do.by
	methods. The function takes as input a dataframe from which a small sample is to be returned. It is also possible to specify the desired sampling
	fraction (second argument). The value of sampling fraction is between 0 and 1. The third argument, \texttt{is.random}, is a boolean value. If TRUE, a random sample is drawn, else it takes the \texttt{head()} of the data.
	
	The \texttt{rhead()} generates small random samples from a vector or a dataframe. The function takes input a vector or a dataframe from which the small random samples are generated. The second argument is a
	positive integer, representing the number of lines or elements to print. The default value is 7. The third argument, \texttt{is.random} is same as the one of the above functions. All functions other than \texttt{CompEst} and \texttt{rhead()} are internal and not directly accessible to users, but corresponding codes are available at Github page \cite{github_main}.
	
	\hypertarget{demonstration-of-guesscompx-package}{%
		\section{\texorpdfstring{Demonstration of \texttt{Guesscompx}
				package}{Demonstration of Guesscompx package}}\label{demonstration-of-guesscompx-package}}
	
	This section demonstrates the usefulness of the \texttt{Guesscompx} package. It provides several examples showing some use cases (distance function, time series, custom function) and how to attain the best tuned parameters. It is important to keep in mind, since this package is measuring
	actual CPU times, absolute reproducibility is out of reach. Running the examples on the computer will return slightly different results each time.
	
	\hypertarget{example-1}{%
		\subsection{Example 1}\label{example-1}}
	
	This example sets up a dummy function that mimics an algorithm with a \emph{O(1)} time complexity and a \emph{O(N.log(N))} memory complexity, both with some random noise. See the warning issued for the time
	complexity : when a constant model is found, it always suggests that the cause might be insufficient running time or insufficient replicates. Also note that a \emph{O(N.log(N))} trend can sometimes be mistaken for a linear trend.
	
	\footnotesize
	\begin{verbatim}
	# Dummy function that mimics a constant time complexity and
	# N.log(N) memory complexity:
	f1 = function(df){
	Sys.sleep(rnorm(1, 0.1, 0.02))
	v = rnorm(n = nrow(df)*log(nrow(df))*(runif(1, 1e3, 1.1e3)))
	}
	out = CompEst(d = mtcars, f = f1, replicates=2, start.size=2, max.time = 1)
	\end{verbatim}
	\normalsize
	
	
	\begin{figure}[H]
		\centering
		\begin{subfigure}[b]{0.8\textwidth}
			\centering
			\includegraphics[width=\textwidth]{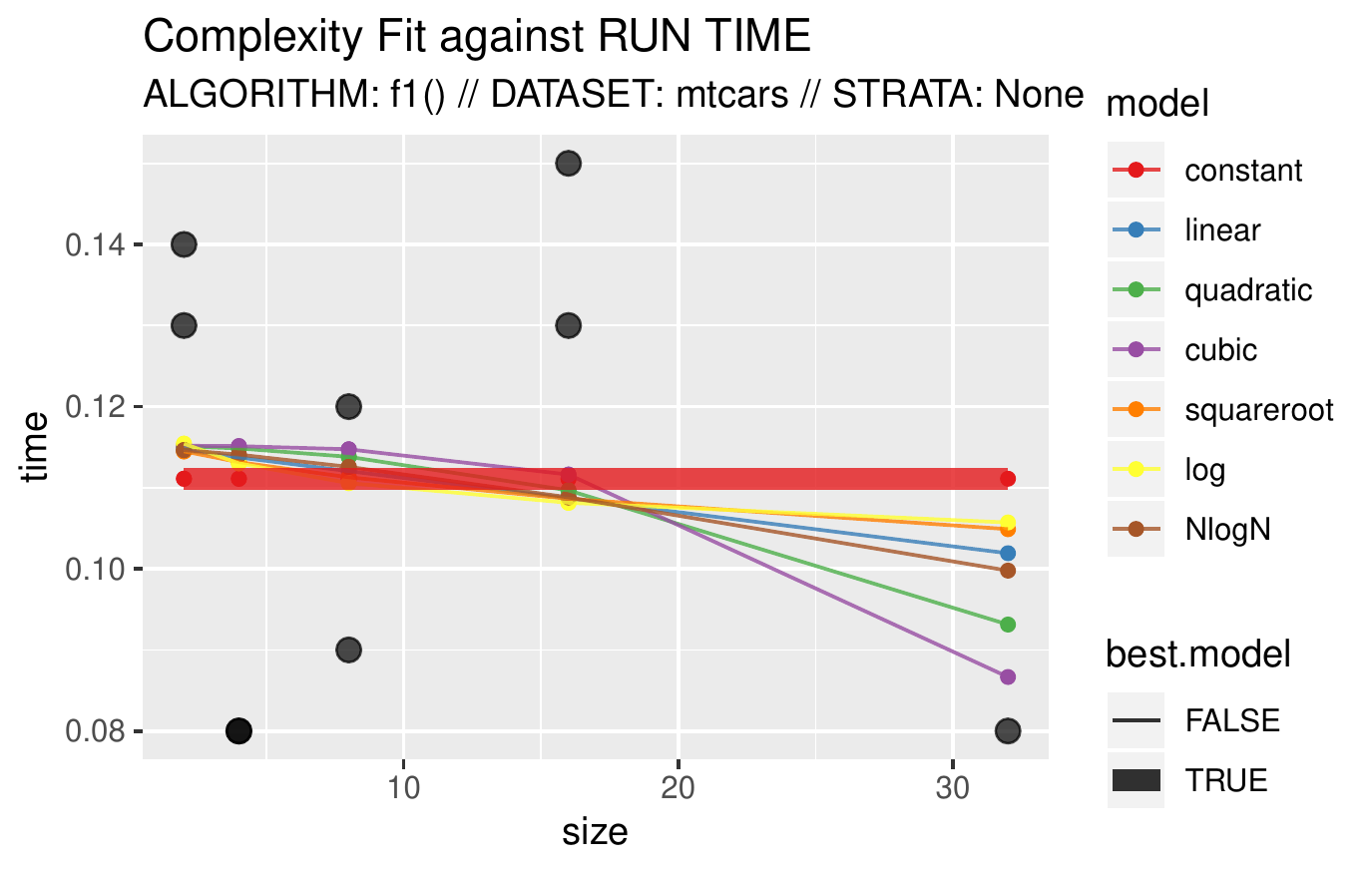}
			\caption{}
			\label{fig:y equals x}
		\end{subfigure}
		\hfill
		\begin{subfigure}[b]{0.8\textwidth}
			\centering
			\includegraphics[width=\textwidth]{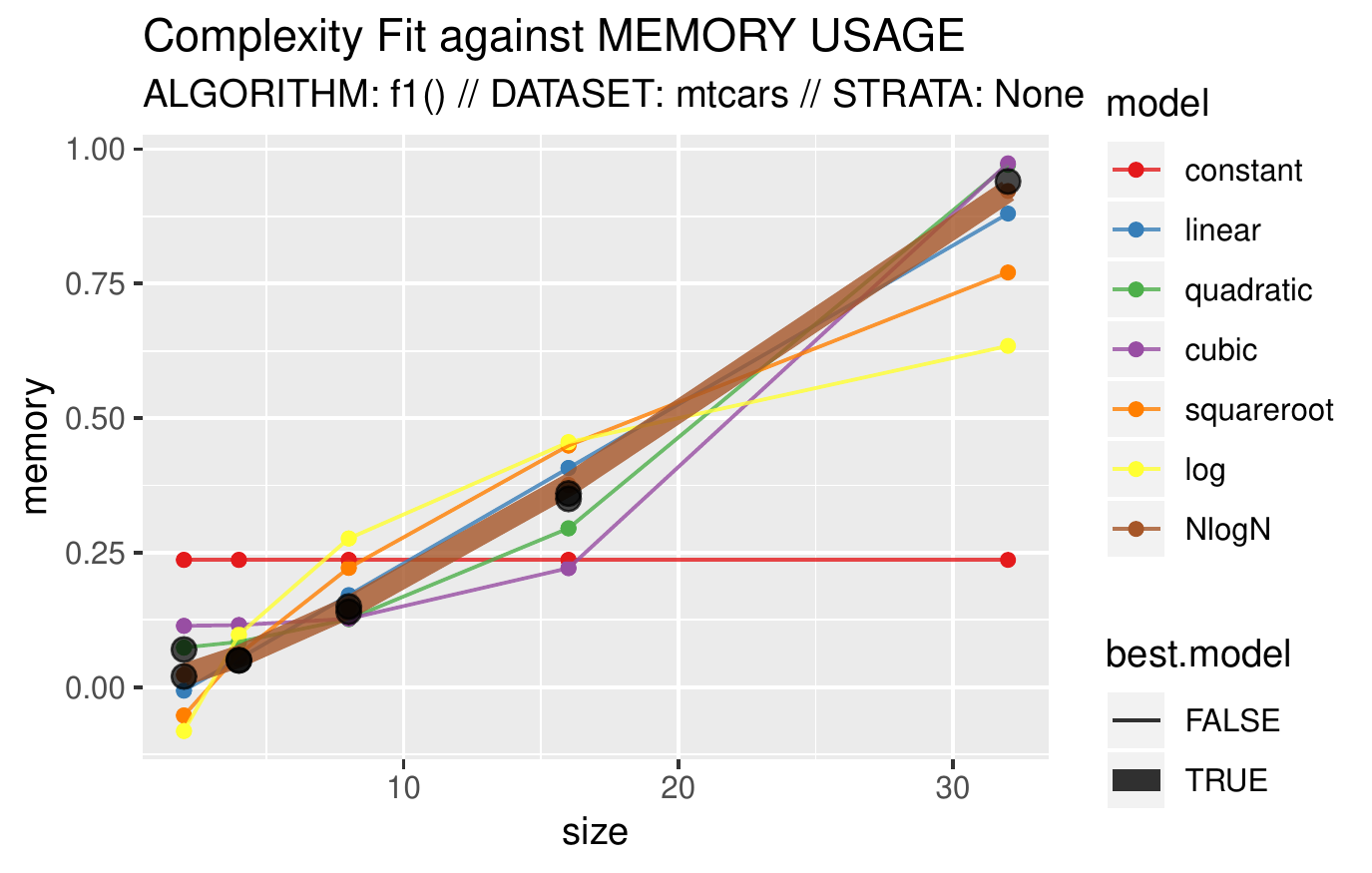}
			\caption{}
			\label{fig:three sin x}
		\end{subfigure}
		\caption{Complexity fit against (a) run time and (b) memory usage for function `f()` on `mtcars` dataset, suggests \emph{O(1)}  and \emph{O(Nlog(N))} as best model, respectively (Example 1)}
		\label{fig:unnamed-chunk-31}
	\end{figure}
	
	\footnotesize	
	\begin{verbatim}
	# Raises an alert for TIME complexity.
	# Sometimes confuses MEMORY complexity with linear:
	print(out)
	\end{verbatim}
	\normalsize
	\footnotesize
	\begin{verbatim}
	#> $sample.sizes
	#>  [1]  2  2  4  4  8  8 16 16 32 32
	#> 
	#> $`TIME COMPLEXITY RESULTS`
	#> $`TIME COMPLEXITY RESULTS`$best.model
	#> [1] "CONSTANT"
	#> 
	#> $`TIME COMPLEXITY RESULTS`$computation.time.on.full.dataset
	#> [1] "0.11S"
	#> 
	#> $`TIME COMPLEXITY RESULTS`$p.value.model.significance
	#> [1] NA
	#> 
	#> 
	#> $`MEMORY COMPLEXITY RESULTS`
	#> $`MEMORY COMPLEXITY RESULTS`$best.model
	#> [1] "NLOGN"
	#> 
	#> $`MEMORY COMPLEXITY RESULTS`$memory.usage.on.full.dataset
	#> [1] "1 Mb"
	#> 
	#> $`MEMORY COMPLEXITY RESULTS`$system.memory.limit
	#> [1] "8064 Mb"
	#> 
	#> $`MEMORY COMPLEXITY RESULTS`$p.value.model.significance
	#> [1] 3.166196e-09
	\end{verbatim}
	\normalsize
	
	\hypertarget{example-2}{%
		\subsection{Example 2}\label{example-2}}
	
	This example tests the behaviour against a real-life distance algorithm
	whose complexity should be a clear \emph{O(N\^{}2)}, for both memory and
	time.
	\footnotesize		
	\begin{verbatim}
	# 'dist' function analysis:
	f2 = dist
	d  = ggplot2::diamonds[, 5:8]
	CompEst(d = d, f = f2, replicates = 1, max.time = 1)
	\end{verbatim}
	\normalsize
	
	
	\begin{figure}[H]
		\centering
		\begin{subfigure}[b]{0.49\textwidth}
			\centering
			\includegraphics[width=\textwidth]{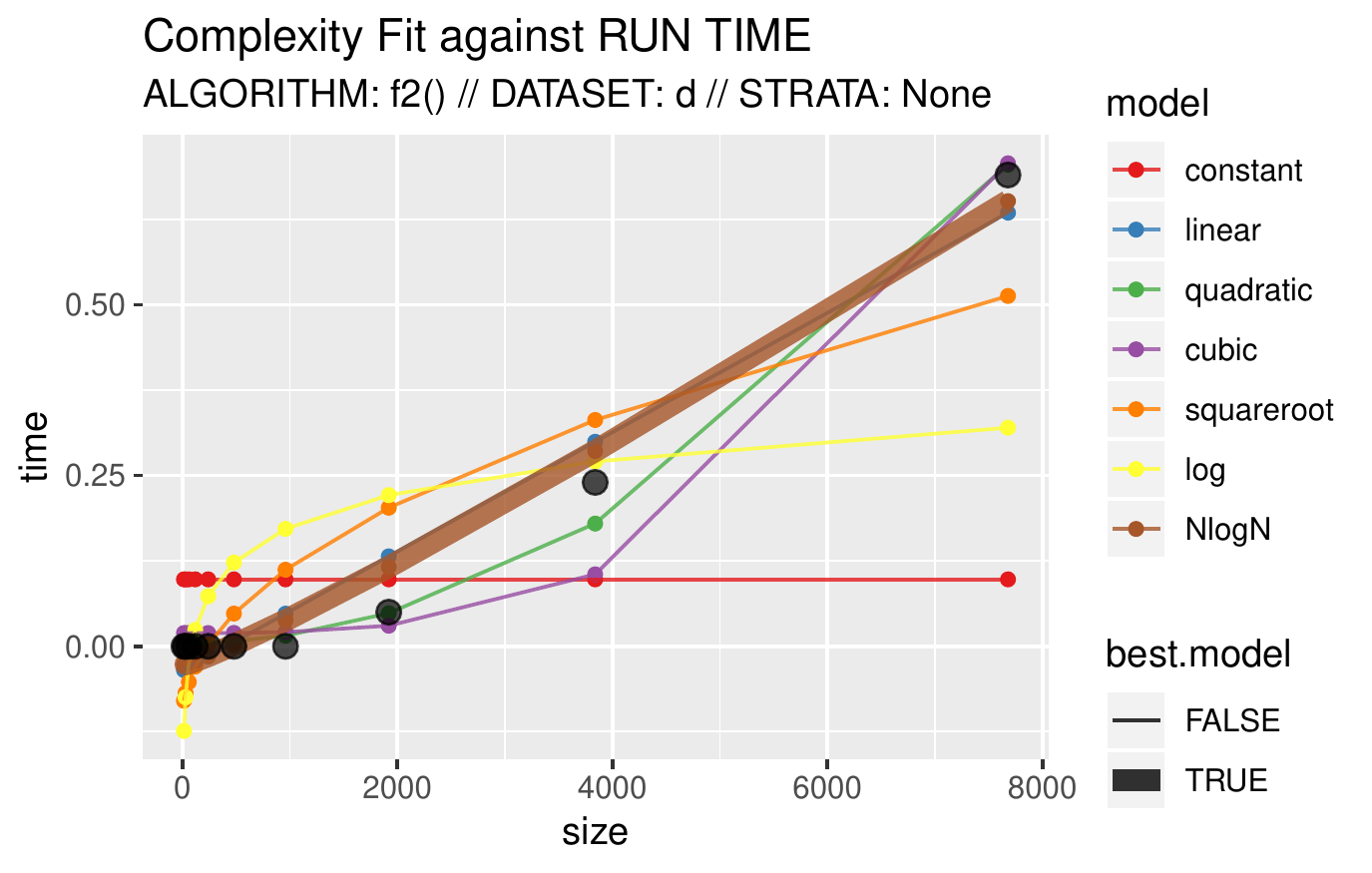}
			\caption{}
			\label{fig:y equals x}
		\end{subfigure}
		\hfill
		\begin{subfigure}[b]{0.49\textwidth}
			\centering
			\includegraphics[width=\textwidth]{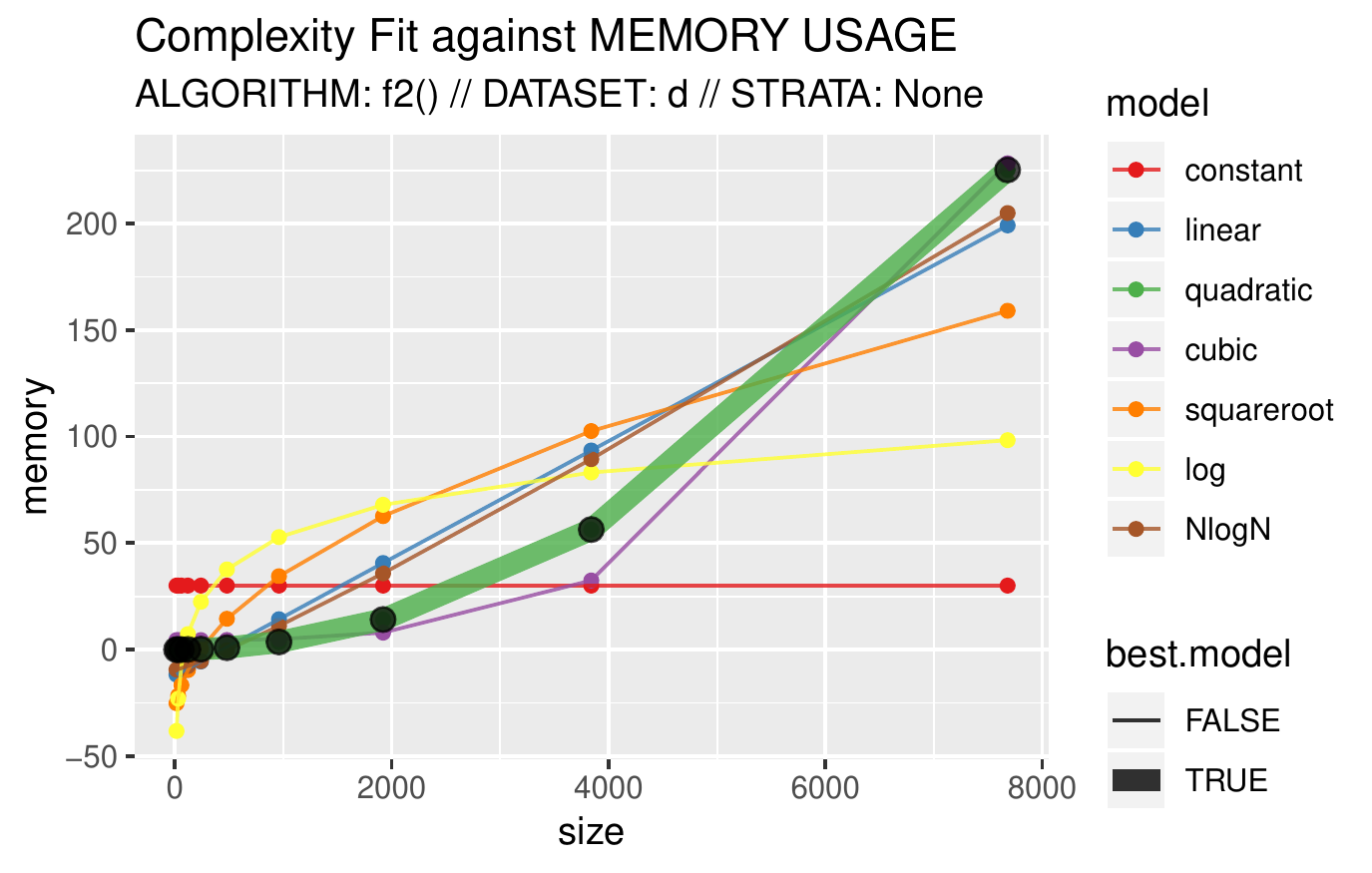}
			\caption{}
			\label{fig:three sin x}
		\end{subfigure}
		\caption{Complexity fit against (a) run time and (b) memory usage for distance function on `diamonds` dataset, suggests \emph{O(Nlog(N))}  and \emph{O(N\^{}2)} as best model, respectively (Example 2)}
		\label{fig:unnamed-chunk-41}
	\end{figure}
	
	\footnotesize			
	\begin{verbatim}
	#> $sample.sizes
	#>  [1]    15    30    60   120   240   480   960  1920  3840  7680 15360
	#> [12] 30720 53940
	#> 
	#> $`TIME COMPLEXITY RESULTS`
	#> $`TIME COMPLEXITY RESULTS`$best.model
	#> [1] "NLOGN"
	#> 
	#> $`TIME COMPLEXITY RESULTS`$computation.time.on.full.dataset
	#> [1] "5.78S"
	#> 
	#> $`TIME COMPLEXITY RESULTS`$p.value.model.significance
	#> [1] 1.57405e-07
	#> 
	#> 
	#> $`MEMORY COMPLEXITY RESULTS`
	#> $`MEMORY COMPLEXITY RESULTS`$best.model
	#> [1] "QUADRATIC"
	#> 
	#> $`MEMORY COMPLEXITY RESULTS`$memory.usage.on.full.dataset
	#> [1] "11110 Mb"
	#> 
	#> $`MEMORY COMPLEXITY RESULTS`$system.memory.limit
	#> [1] "8064 Mb"
	#> 
	#> $`MEMORY COMPLEXITY RESULTS`$p.value.model.significance
	#> [1] 1.28452e-28
	\end{verbatim}
	\normalsize
	
	\hypertarget{example-3}{%
		\subsection{Example 3}\label{example-3}}
	
	This example tests the time and memory complexity of a time series prediction method. For this purpose, an ARIMA model is used. For time series functions, the \texttt{f} argument may include \texttt{ts()} and
	to avoid loosing this \texttt{ts} attribute at sampling, it is recommended to set \texttt{start.size} argument to 3 periods at least. The ARIMA function should return a linear trend for time complexity.
	
	\footnotesize				
	\begin{verbatim}
	# time series prediction function analysis:
	f = function(d) arima(ts(d, freq = 12), order=c(1,0,1), seasonal = c(0,1,1))
	d = ggplot2::txhousing$sales
	CompEst(d, f, start.size = 4*12, random.sampling = FALSE)
	\end{verbatim}
	\normalsize
	

	\begin{figure}[H]
		\centering
		\begin{subfigure}[b]{0.49\textwidth}
			\centering
			\includegraphics[width=\textwidth]{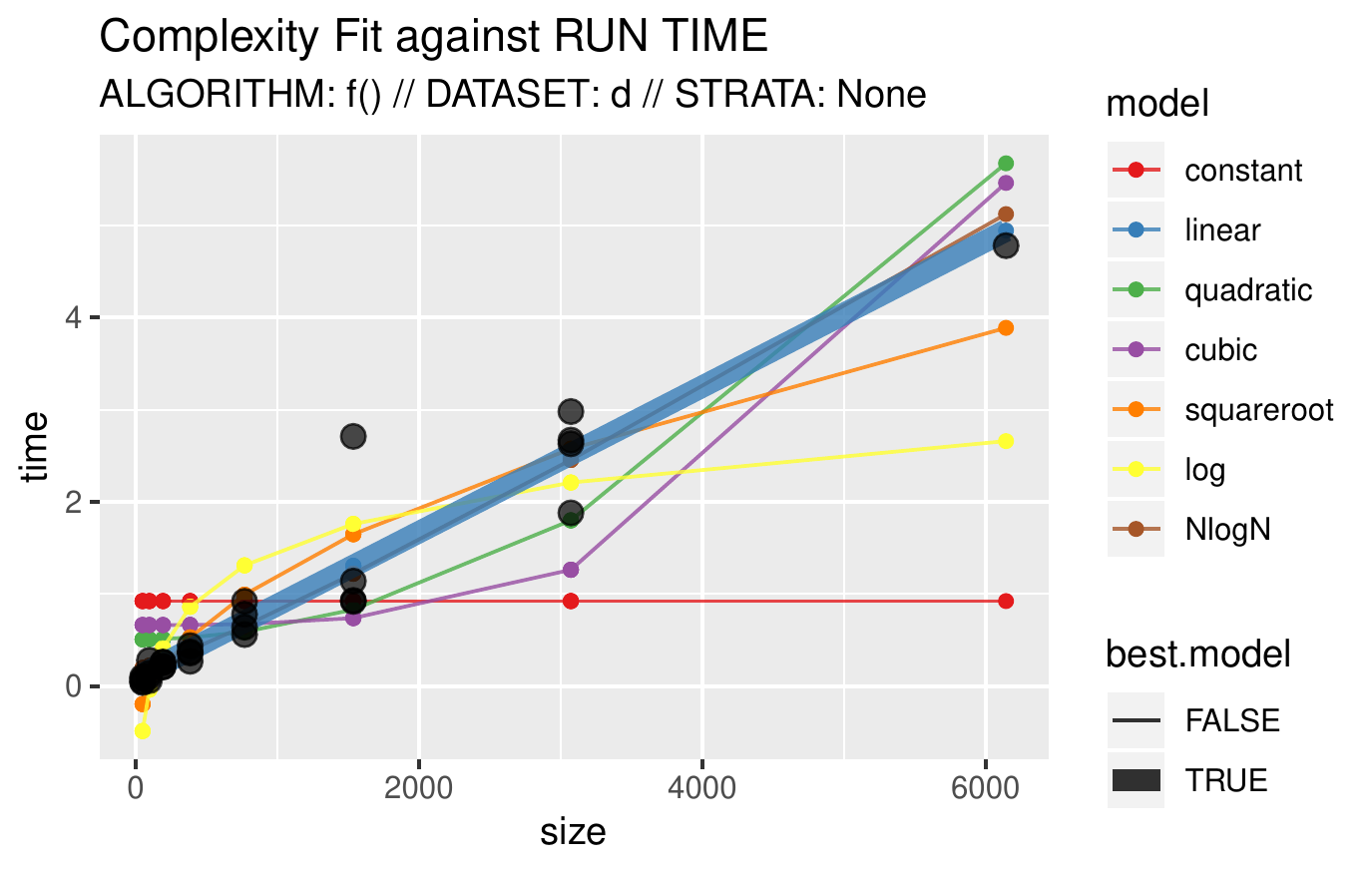}
			\caption{}
			\label{fig:y equals x}
		\end{subfigure}
		\hfill
		\begin{subfigure}[b]{0.49\textwidth}
			\centering
			\includegraphics[width=\textwidth]{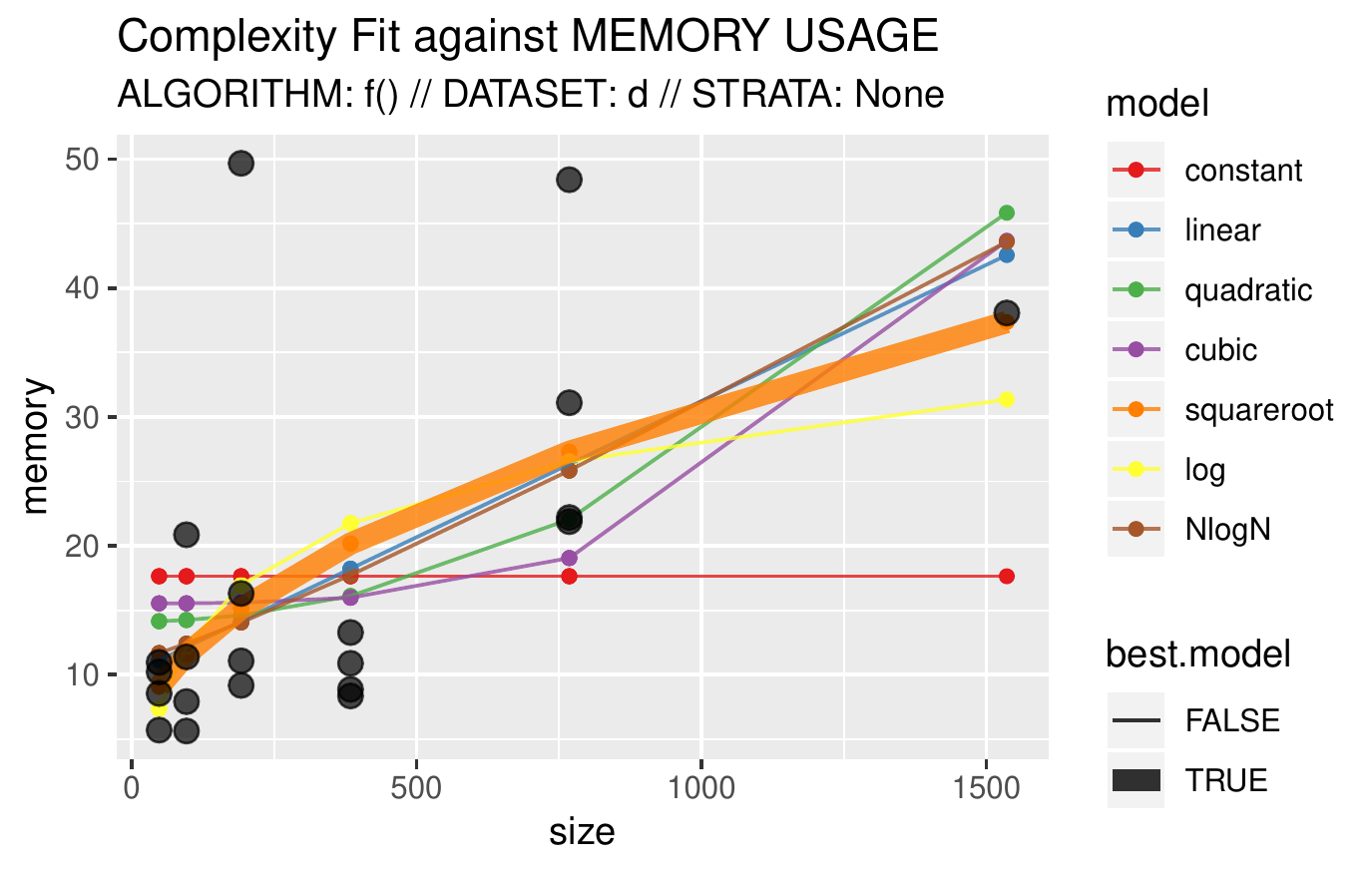}
			\caption{}
			\label{fig:three sin x}
		\end{subfigure}
		\caption{Complexity fit against (a) run time and (b) memory usage for distance function on `sales` dataset, suggests \emph{O(N)}  and \emph{O(N\^{}(1/2))} as best model, respectively (Example 3)}
		\label{fig:unnamed-chunk-51}
	\end{figure}
	
	\footnotesize						
	\begin{verbatim}
	#> $sample.sizes
	#>  [1]   48   48   48   48   96   96   96   96  192  192  192  192  384  384
	#> [15]  384  384  768  768  768  768 1536 1536 1536 1536 3072 3072 3072 3072
	#> [29] 6144 6144 6144 6144 8602 8602 8602 8602
	#> 
	#> $`TIME COMPLEXITY RESULTS`
	#> $`TIME COMPLEXITY RESULTS`$best.model
	#> [1] "LINEAR"
	#> 
	#> $`TIME COMPLEXITY RESULTS`$computation.time.on.full.dataset
	#> [1] "6.88S"
	#> 
	#> $`TIME COMPLEXITY RESULTS`$p.value.model.significance
	#> [1] 3.492077e-16
	#> 
	#>  
	#> $`MEMORY COMPLEXITY RESULTS`
	#> $`MEMORY COMPLEXITY RESULTS`$best.model
	#> [1] "SQUAREROOT"
	#> 
	#> $`MEMORY COMPLEXITY RESULTS`$memory.usage.on.full.dataset
	#> [1] "84 Mb"
	#> 
	#> $`MEMORY COMPLEXITY RESULTS`$system.memory.limit
	#> [1] "8064 Mb"
	#> 
	#> $`MEMORY COMPLEXITY RESULTS`$p.value.model.significance
	#> [1] 0.004649934
	\end{verbatim}
	\normalsize							
	\hypertarget{example-4}{%
		\subsection{Example 4}\label{example-4}}
	
	This example tests the behavior of eigen decomposition process of a matrix. Since the algorithm only works on square matrices, here the target function subsets columns to force the matrix into a square one,
	otherwise the sampling would result in a rectangular matrix. This analysis yields an \emph{O(N\^{}3)} fit for time complexity, which is scientifically correct, and a quadratic for memory one.
	
	\footnotesize
	\begin{verbatim}
	# Eigendecomposition of a matrix
	m = matrix(rnorm(1e6), ncol=1000, nrow=1000)
	
	# force the matrix into a square one:
	eigen. = function(m) eigen(as.matrix(m[, 1:nrow(m)]))
	
	# This yields an O(n^3) fit, which is scientifically correct, 
	# and a quadratic for memory. 
	out = CompEst(m, eigen., replicates = 5, max.time = 60)
	\end{verbatim}
	\normalsize
	

	\begin{figure}[H]
		\centering
		\begin{subfigure}[b]{0.49\textwidth}
			\centering
			\includegraphics[width=\textwidth]{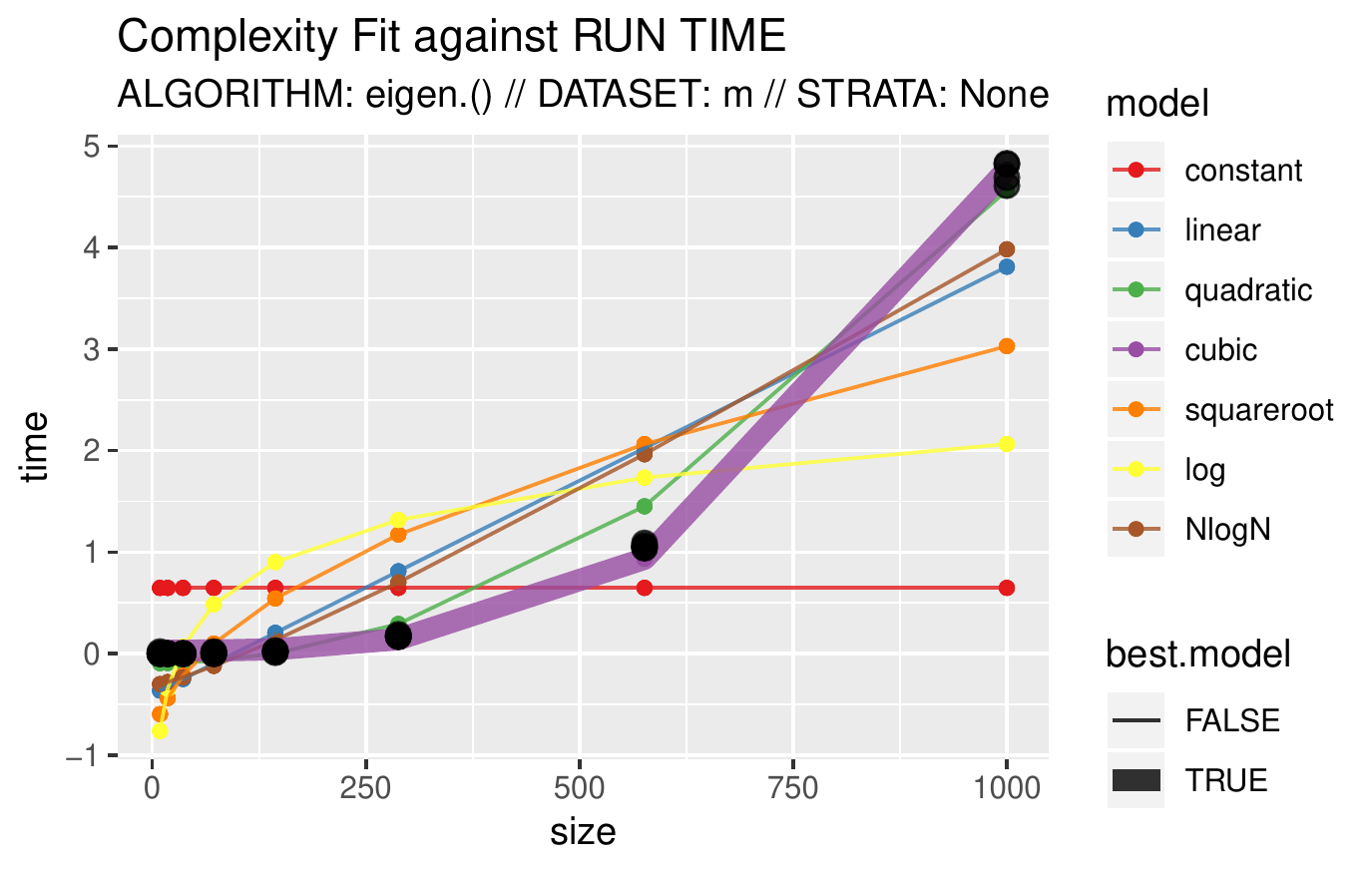}
			\caption{}
			\label{fig:y equals x}
		\end{subfigure}
		\hfill
		\begin{subfigure}[b]{0.49\textwidth}
			\centering
			\includegraphics[width=\textwidth]{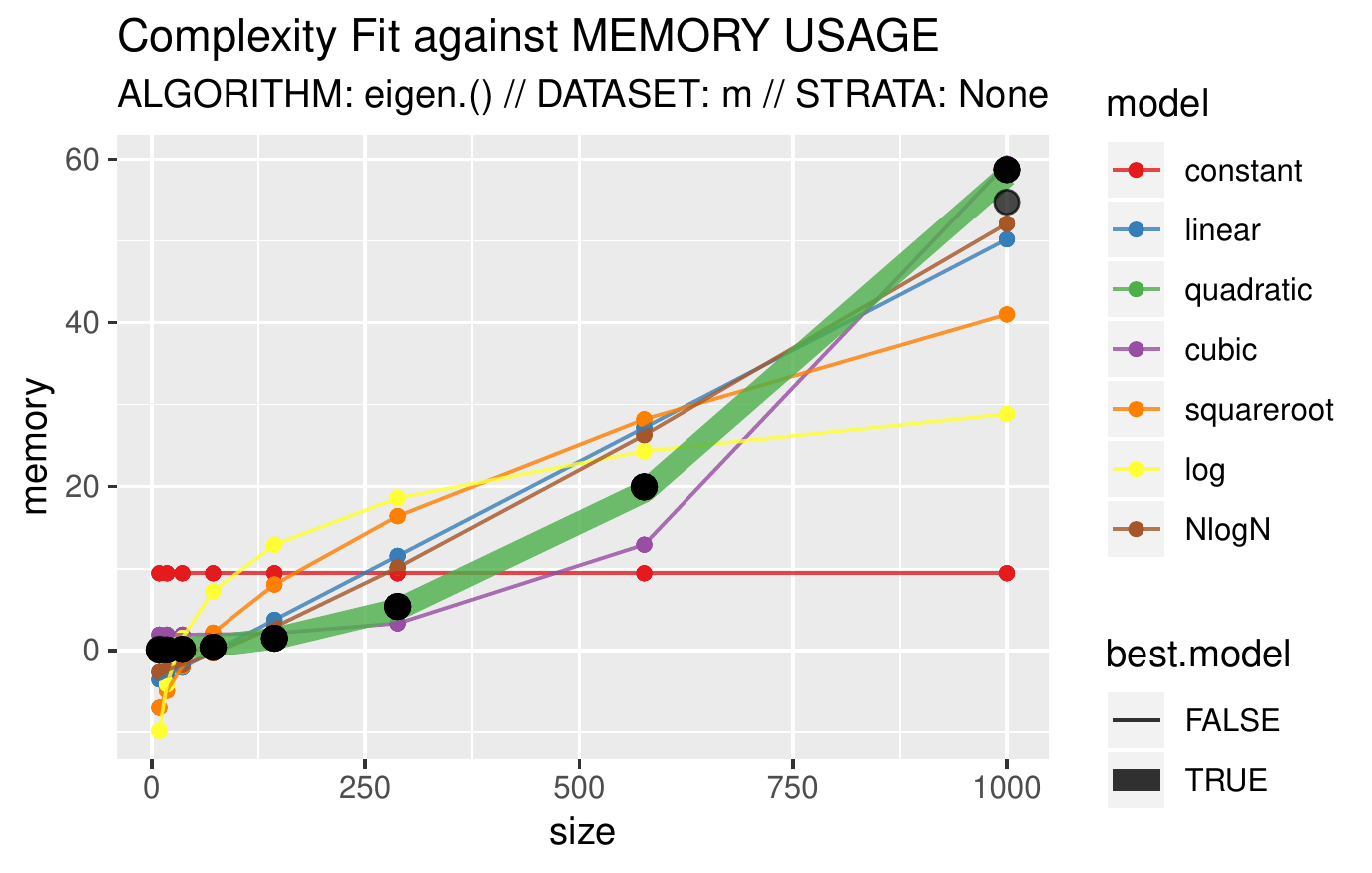}
			\caption{}
			\label{fig:three sin x}
		\end{subfigure}
		\caption{Complexity fit against (a) run time and (b) memory usage for eigen decomposition function on a matrix dataset, suggests \emph{O(N\^{}3)}  and \emph{O(N\^{}2)} as best model, respectively (Example 4)}
		\label{fig:unnamed-chunk-61}
	\end{figure}
	
	\hypertarget{example-5}{%
		\subsection{Example 5}\label{example-5}}
	
	This example is an special case of stratified input, which is useful to force the sampling to have at least one observation of each class of a specific column. Consider a function to predict a diabetes outcome in
	the dataset `PimaIndiansDiabetes` with the Support Vector Machine (SVM) method. Depending on algorithm, the true complexity (time) of SVM is between \emph{O(N\^{}2)} and \emph{O(N\^{}3)}. 
	
	\footnotesize
	\begin{verbatim}
	library(mlbench)
	library(e1071)
	data("PimaIndiansDiabetes")
	f6 = function(df){
	fit = svm(diabetes ~ ., data=df)
	return(table(df$diabetes, fitted(fit)))
	}
	\end{verbatim}
	\normalsize
	
	With this configuration, sampling will often result in only negative or positive observations to train the model, which results in an error message as shown below.
	
	\footnotesize
	\begin{verbatim}
	CompEst(PimaIndiansDiabetes, f6, start.size = 3, power.factor = 3)
	
	# ERROR: model is empty!
	\end{verbatim}
	\normalsize
	
	After fixing the strata argument to be the Y vector of classes, to ensure that both classes will be represented in the model as follows. Currently, the strata argument enables to specify only one column, but
	future versions of this package will accept list of columns.
	
	\footnotesize
	\begin{verbatim}
	CompEst(PimaIndiansDiabetes, f6, start.size = 3, power.factor = 3, strata = "diabetes")
	\end{verbatim}
	\normalsize


	\begin{figure}[H]
		\centering
		\begin{subfigure}[b]{0.49\textwidth}
			\centering
			\includegraphics[width=\textwidth]{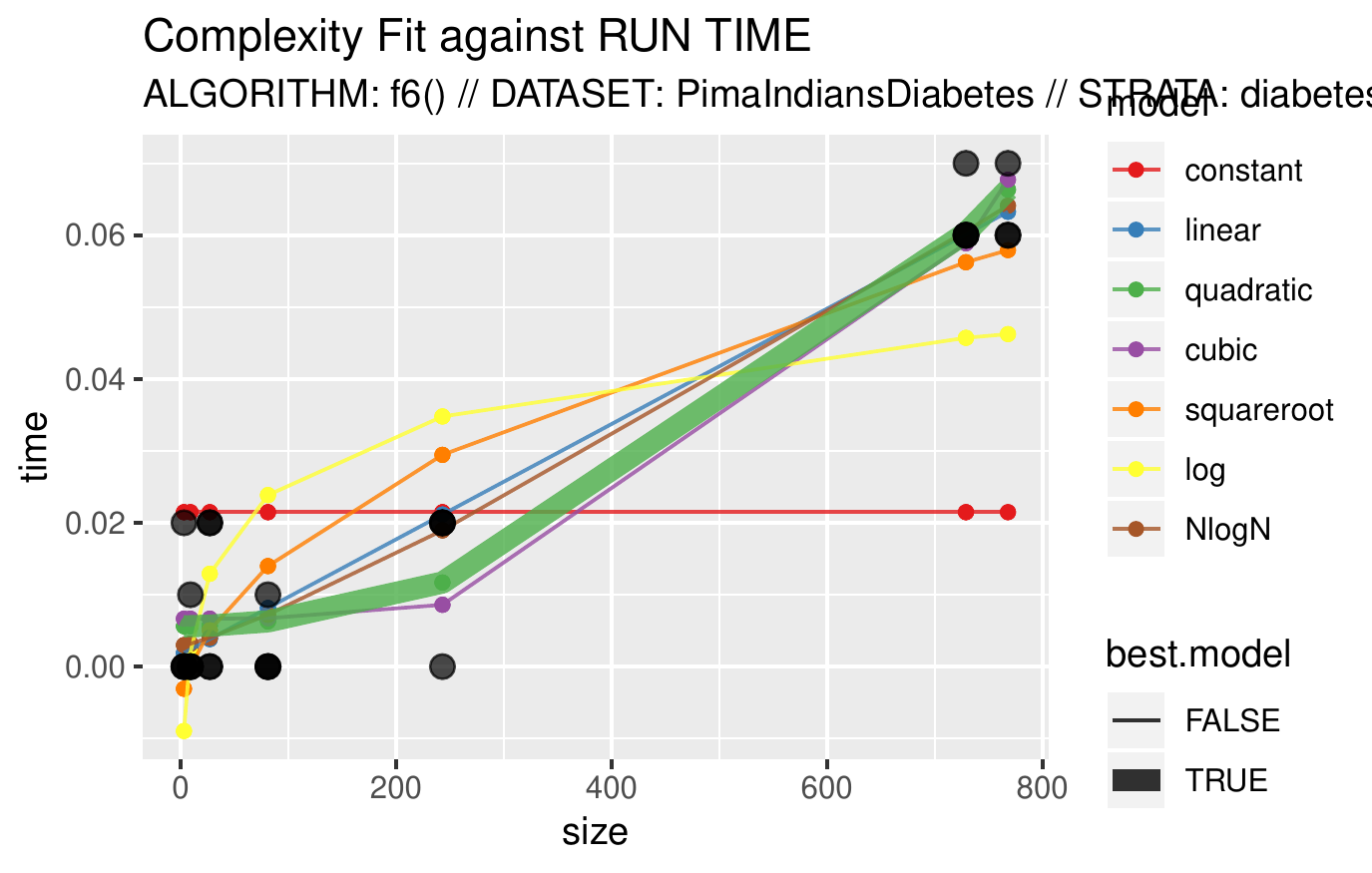}
			\caption{}
			\label{fig:y equals x}
		\end{subfigure}
		\hfill
		\begin{subfigure}[b]{0.49\textwidth}
			\centering
			\includegraphics[width=\textwidth]{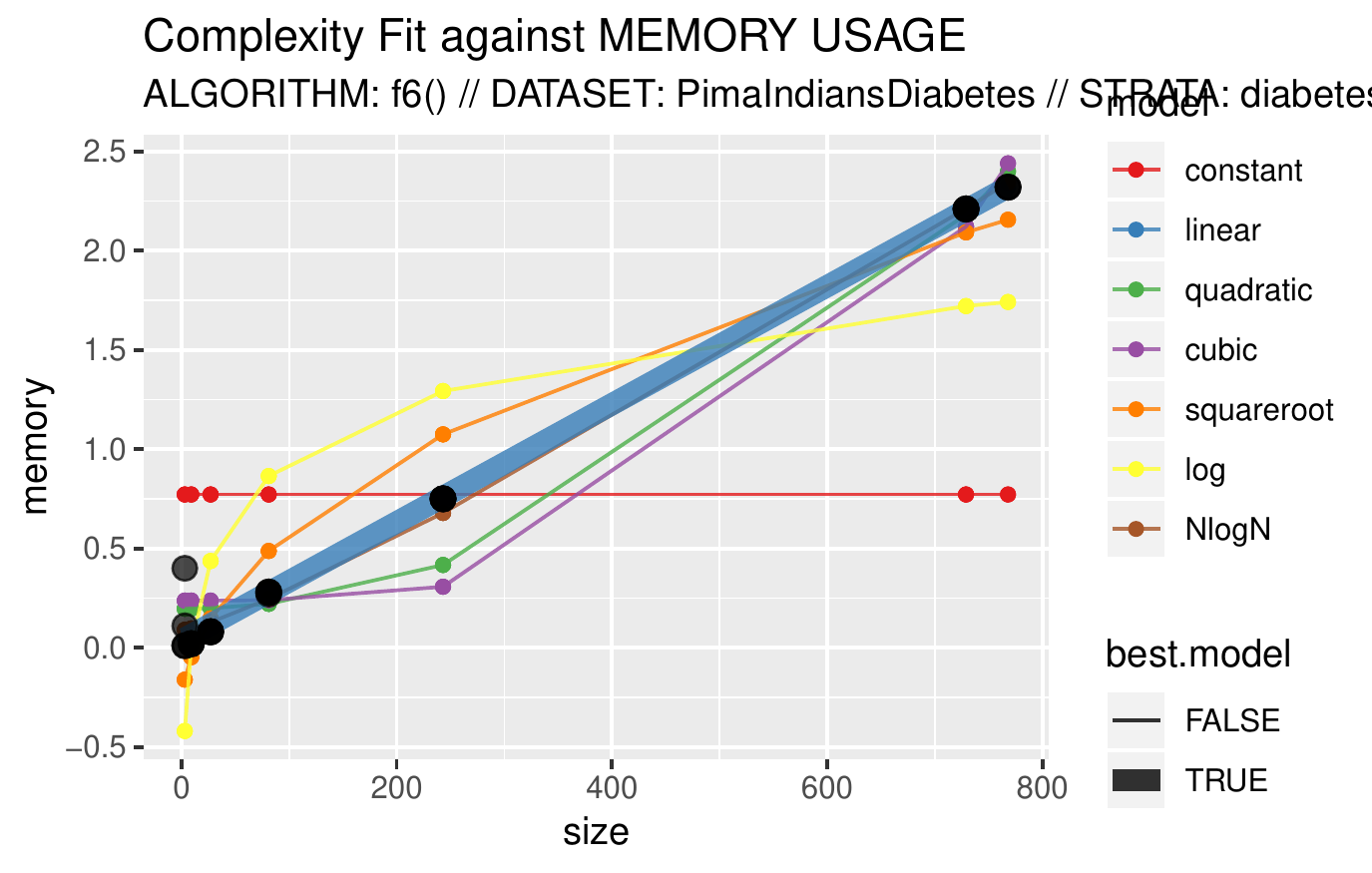}
			\caption{}
			\label{fig:three sin x}
		\end{subfigure}
		\caption{Complexity fit against (a) run time and (b) memory usage for SVM function on `PimaIndiansDiabetes` dataset, suggests \emph{O(N\^{}2)}  and \emph{O(N)} as best model, respectively (Example 5)}
		\label{fig:unnamed-chunk-91}
	\end{figure}
	
	\footnotesize
	\begin{verbatim}
	#> $sample.sizes
	#>  [1]   3   3   3   3   9   9   9   9  27  27  27  27  81  81  81  81 243
	#> [18] 243 243 243 729 729 729 729 768 768 768 768
	#> 
	#> $`TIME COMPLEXITY RESULTS`
	#> $`TIME COMPLEXITY RESULTS`$best.model
	#> [1] "QUADRATIC"
	#> 
	#> $`TIME COMPLEXITY RESULTS`$computation.time.on.full.dataset
	#> [1] "0.07S"
	#> 
	#> $`TIME COMPLEXITY RESULTS`$p.value.model.significance
	#> [1] 8.624223e-15
	#> 
	#> 
	#> $`MEMORY COMPLEXITY RESULTS`
	#> $`MEMORY COMPLEXITY RESULTS`$best.model
	#> [1] "LINEAR"
	#> 
	#> $`MEMORY COMPLEXITY RESULTS`$memory.usage.on.full.dataset
	#> [1] "2 Mb"
	#> 
	#> $`MEMORY COMPLEXITY RESULTS`$system.memory.limit
	#> [1] "8064 Mb"
	#> 
	#> $`MEMORY COMPLEXITY RESULTS`$p.value.model.significance
	#> [1] 7.60052e-29
	\end{verbatim}
	\normalsize

	\hypertarget{example-6}{%
		\subsection{Example 6}\label{example-6}}
	
	This example tests the complexity of a dummy function that mimics a linear time complexity with high CPU variability and heteroscedasticity. This function results in a variety of non-significant ``best models'',
	often ``LINEAR'' one. Thereafter, we increase the number of replicates, start at a smaller size, and get an almost always ``Linear'' response. This example takes much more time because we replicate 10
	times the result.
	
	\footnotesize
	\begin{verbatim}
	# A dummy function:
	f3 = function(df){  Sys.sleep(max(0, rnorm(1, 0.1+nrow(df)/500, nrow(df)/2000)))  }
	set.seed(1); 
	replicate(10, CompEst(d = mtcars, f = f3, plot.result = F)$`TIME COMPLEXITY RESULTS`$best.model)
	\end{verbatim}
	\normalsize
	\footnotesize
	\begin{verbatim}
	#>  [1] "LOG"        "SQUAREROOT" "LINEAR"     "LOG"        "SQUAREROOT"
	#>  [6] "QUADRATIC"  "SQUAREROOT" "QUADRATIC"  "SQUAREROOT" "SQUAREROOT"
	\end{verbatim}
	\normalsize
	\footnotesize
	\begin{verbatim}
	set.seed(2); 
	replicate(10, CompEst(d = mtcars, f = f3, plot.result = F, start.size = 1, 
	replicates = 10, max.time=5)$`TIME COMPLEXITY RESULTS`$best.model)
	\end{verbatim}
	\normalsize
	\footnotesize
	\begin{verbatim}
	#>  [1] "SQUAREROOT" "LINEAR"     "NLOGN"      "SQUAREROOT" "LINEAR"    
	#>  [6] "LINEAR"     "LINEAR"     "LINEAR"     "NLOGN"      "NLOGN"
	\end{verbatim}
	\normalsize

	\hypertarget{performance}{%
		\section{Performance}\label{performance}}
	
	As in \cite{kumarsharma2018predicting}, it was assessed the prediction power of the described method. It is important to understand the prediction performance of the proposed tool has no absolute meaning, since it can always be improved by allocating more time to run the test.

	\begin{table}[h]
		\begin{minipage}{\linewidth}
			\centering
			\caption{Some famous algorithms with their predicted complexities}
			\label{t1} {
				\begin{tabular}{lllll} 
					\toprule
					\textbf{} & \begin{tabular}[c]{@{}l@{}}True time \\ complexity\end{tabular} & \begin{tabular}[c]{@{}l@{}}Time \\ Accuracy\end{tabular} & \begin{tabular}[c]{@{}l@{}}True space\\ complexity\end{tabular} & \begin{tabular}[c]{@{}l@{}}Predicted space\\ complexity\end{tabular} \\ \midrule
					Bubble sort & O(N\textasciicircum{}2) & 97\% & O(N) & 97\% O(Nlog(N))\footnotemark[1] \\
					Find max & O(N) & 95\% & O(1) & 65\%  \\
					Permutations & O(N\textasciicircum{}3) & 99\% & O(N) & 95\%  \\
					Tree split\footnotemark[2] & O(log(N)) & 38\% & O(NlogN) & 98\%  \\
					Shell search & O(N\textasciicircum{}(4/3))\footnotemark[3] & \begin{tabular}[c]{@{}l@{}}NA \\ (73\% O(NlogN), \\ 27\% O(N))\end{tabular} & O(N) & 100\%  \\ \bottomrule
			\end{tabular}}\\
			\footnotemark[1]{it is known that \emph{O(logN)} and \emph{O(NlogN)} are easy to confuse}\\
			\footnotemark[2]{rpart regression from rpart package, which is sensitive to the data itself}\\
			\footnotemark[3]{not part of the 7 base complexity functions}
		\end{minipage}
	\end{table}

	These considerations taken aside, a test bench was set up in order to test one paragon algorithm for each complexity family as shown in Table \ref{t1}. {The selected algorithms are the popular ones with known time and space complexities. In this section, we have checked the performance of the \texttt{GuessCompx} in predicting the respective time and space complexities.} Whenever possible, we worked with the default setting of the \texttt{CompEst()} function, sometimes reducing the time limit down to 1
	second. One hundred replicates of the function were set. Accuracy is defined as:
	\begin{equation}
	\text{Accuracy} = \frac{\text{Number of correct predictions for complexity function}}{\text{total number of cases}} \times 100
	\end{equation}
	
	{Table \ref{t1} shows the accuracy in predicting complexities of the targeted algorithms. In most of the cases (especially, Bubble sort, Find max, Permutation), both predicted time and space complexities are significantly accurate. Further, the space complexity of the Tree split algorithm is predicted accurately, but time complexity accuracy is average, since the algorithm is highly sensitive to the input dataset. Besides, the true time complexity of the Shell search algorithm is O(N\textasciicircum{}(4/3)), but it is not within the seven base complexities considered in the \texttt{GuessCompx} package. Therefore, the predicted time complexities are the ones which are near to the real one. However, the predicted space complexity for the Shell search is 100\% accurate.}
	
\begin{figure}[H]
		\includegraphics{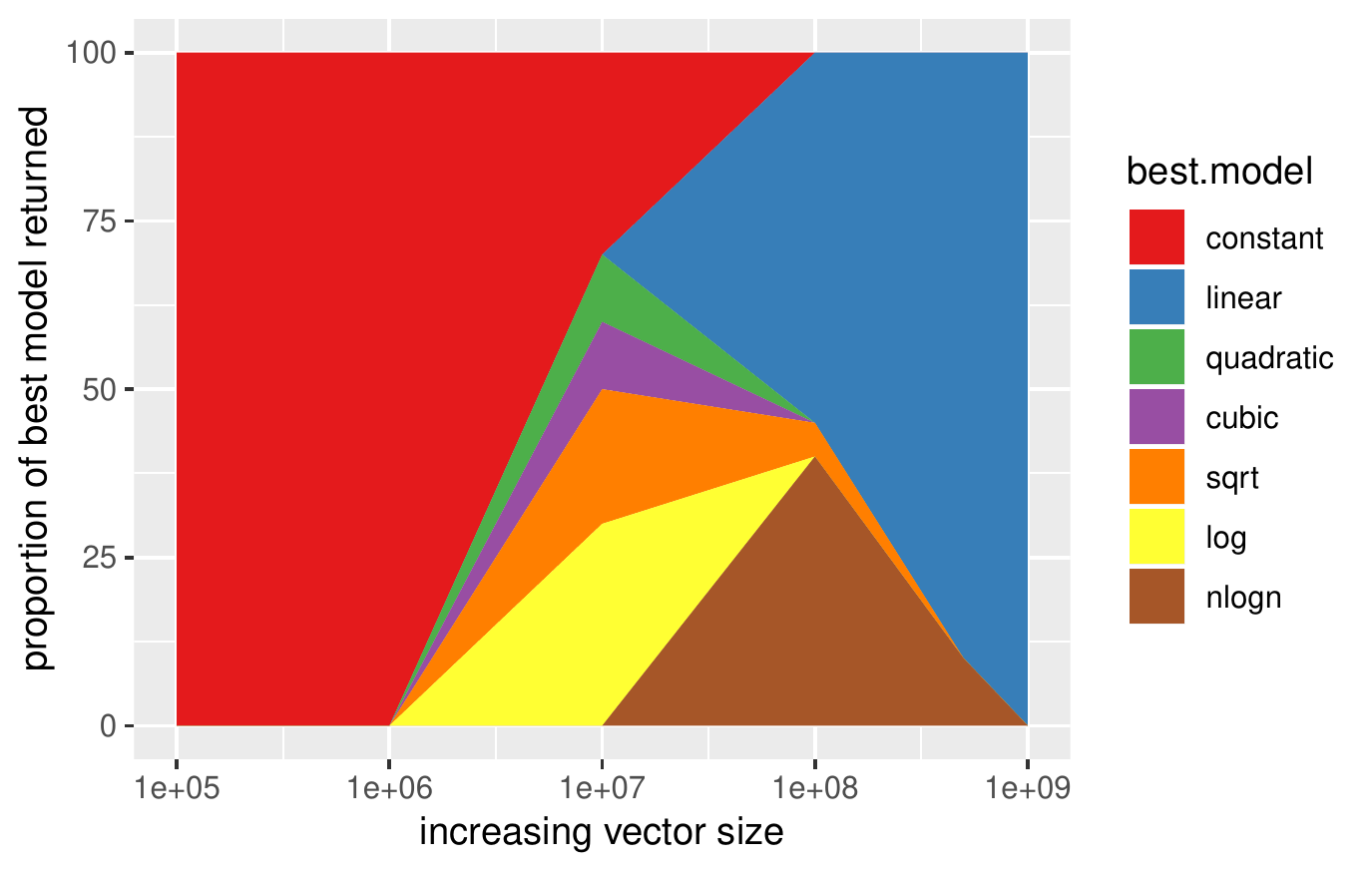} \caption[The asymptotic linear trend for  `max()` function as target algorithm]{The asymptotic linear trend for  `max()` function as target algorithm}\label{fig:unnamed-chunk-10}
	\end{figure}

	Eventually, some of the precision results may seem disappointing. There is indeed a strong sensibility to the choice of the settings and fine-tuning the arguments can make a sensible difference in the result. It is illustrated by a detailed simulation where the maximum vector length varies between 1E4 and 1E9, this last value being considered as the
	''asymptotic'' value. The Figure \ref{fig:unnamed-chunk-10} shows asymptotic linear trend for  `max()` function as target algorithm. It represents the result in proportion of the best model outputted: when the data is too small no model can be determined because the `max()` function is too fast; for vectors around size 1E07, CPU time starts to rise with some variability, but the fit will not return a stable solution; eventually, with sizes going over 1E08, the asymptotic linear trend appears strongly.

	\hypertarget{number-of-replicates}{%
		\subsection{Number of Replicates:}\label{number-of-replicates}}
	
\begin{figure}[H]
		\centering
		\includegraphics[scale=0.7]{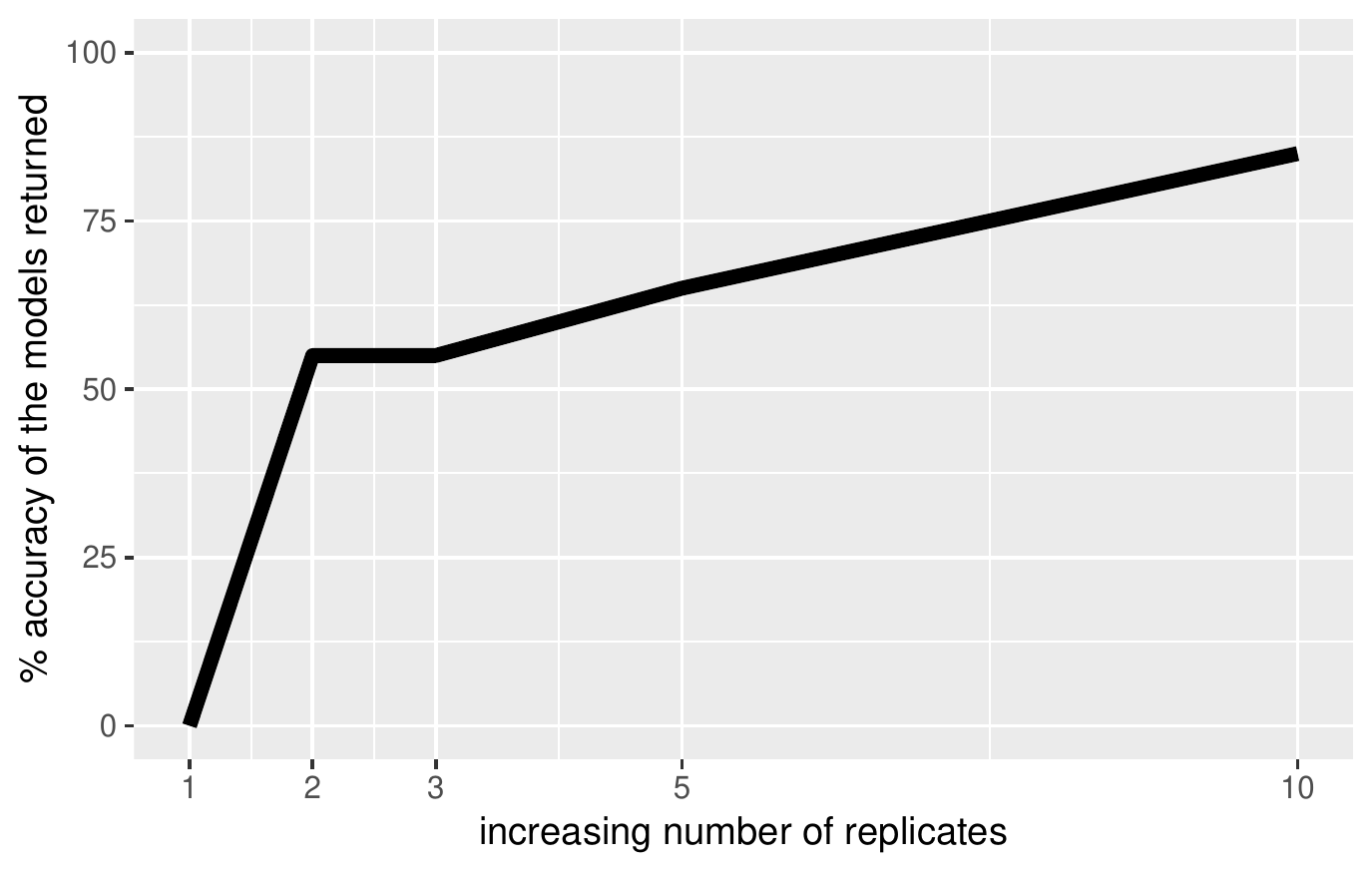} \caption[Effect of number of replicates on finding the correct model]{Effect of number of replicates on finding the correct model for distance algorithm on `diamonds` dataset}\label{fig:unnamed-chunk-11}
	\end{figure}

	As a second performance example, the number of replicates is tested in the benchmark. The \texttt{max.time} limit argument is not used here because it is counted for all replicates of the same size. In order to hold everything constant but the number of replicates, we set an infinite time limit for
	a run, only limiting it with a small input dataset, so it will  not take more than a fraction of seconds to complete a run. This testbench, using a distance computation on a data.frame as the target algorithm, shows that when the we are in the ``grey zone'' (uncertain outcome of the analysis), the number of replicates can make the difference to find the correct model as shown in Figure \ref{fig:unnamed-chunk-11}.

	The codes to test all analysis and comparison done in this section are available in GitHub page \cite{github}.
	
	\hypertarget{summary}{%
		\section{Summary}\label{summary}}
	
	This article was proposed a thorough description of the \texttt{GuessCompx} R package. The package is introduced to perform an empirical estimate on the time and memory complexities of an algorithm or a function. It tests multiple, increasing-sizes samples of the user's data and try to fit one of seven complexity functions: \emph{O(N), O(N\^{}2), O(log(N))}, etc. Based on the best fit procedure using LOO-MSE (leave one out-mean square error), it also predicts the full computation time and memory usage on the whole dataset. {The hereby suggested method and package are believed to be new to the R users community; however there is a lot of room for improvement, both in terms of automation and variety of complexity functions. Further, it is worthy in noting that the complexity is understood only with regard to the size of the data (number of rows), not other possible parameters such as number of features, tuning parameters, etc. Interactions with those parameters could be investigated in future versions of the package. There is always a scope of addition of extra and more complex complexity families in the present version of the \texttt{GuessCompx} package.}


	%
	%

	\bibliographystyle{ieeetr}
	\bibliography{RJreferences}   
	
	
\end{document}